\RequirePackage{amsmath}

\documentclass[runningheads]{llncs}
\usepackage[T1]{fontenc}
\usepackage{graphicx}
\usepackage[hyphens]{url}
\usepackage[hidelinks]{hyperref}
\hypersetup{breaklinks=true}
\usepackage{color}

\usepackage{booktabs}       
\usepackage{csvsimple}
\usepackage{amssymb}
\usepackage{caption}
\usepackage[list=true]{subcaption}

\begin{document}

\title{%
A sandbox study proposal for private and distributed health data analysis
}
\titlerunning{A sandbox study proposal for private and distributed health data analysis}
%
\iftrue
\author{
Rickard Brannvall\inst{}
\and
Hanna Svensson\inst{}
\and
Kannaki Kaliyaperumal
\and \\
Håkan Burden\inst{}
\and
Susanne Stenberg\inst{}
}
\fi
%
\authorrunning{R. Brannvall et al.}
%
\institute{Digital Systems Division, RISE Research Institutes of Sweden 
\email{\{firstname.lastname\}@ri.se}
}
\maketitle              
\begin{abstract}
This paper presents a sandbox study proposal focused on the distributed processing of personal health data within the Vinnova-funded SARDIN project. The project aims to develop the Health Data Bank (Hälsodatabanken in Swedish), a secure platform for research and innovation that complies with the European Health Data Space (EHDS) legislation. By minimizing the sharing and storage of personal data, the platform sends analysis tasks directly to the original data locations, avoiding centralization. This approach raises questions about data controller responsibilities in distributed environments and the anonymization status of aggregated statistical results. The study explores federated analysis, secure multi-party aggregation, and differential privacy techniques, informed by real-world examples from clinical research on Parkinson’s disease, stroke rehabilitation, and wound analysis. To validate the proposed study, numerical experiments were conducted using four open-source datasets to assess the feasibility and effectiveness of the proposed methods. The results support the methods for the proposed sandbox study by demonstrating that differential privacy in combination with secure aggregation techniques significantly improves the privacy-utility trade-off. 
\end{abstract}

\section{Introduction}

This manuscript outlines a sandbox study proposal that could be carried out in the context of regulatory sandboxes as recently introduced both at the EU level \cite{EU_AI_act_regulatory_sandboxes} and at the national level in Sweden \cite{imy2024sandbox}. 
It builds on the use of privacy-enhancing technologies in the SARDIN project \cite{rise2024sardin} for distributed health data analysis, 
which builds a platform called Health Data Bank. This platform's purpose is to utilize healthcare data for research and innovation securely. Health Data Bank is designed to comply with the European Health Data Space (EHDS) legislation \cite{ehds2024} regarding the secondary use of health data.

\begin{figure}
\centering
\includegraphics[width=\textwidth]{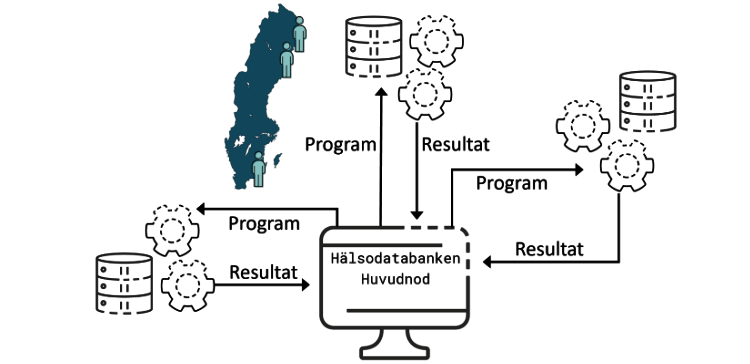}
\caption{Illustration of a network of three edge nodes and a central coordinating node that make up a minimal example of the Health Data Bank. Only program code and aggregate results are communicated over the network, while source data remains at the edge node. Data protection is enhanced by secure aggregation and differential privacy.}
\label{fig:example}
\end{figure}

The platform draws on proven, scalable technology and can be compared to interconnected health data lakes. Each lake represents different health data sources, such as electronic medical records (EMR) from various regions and clinical quality registry data. This allows researchers to analyze sensitive personal data from different data controllers seamlessly. The platform is designed to minimize the sharing and storage of personal data by sending analysis tasks to the original locations where health data is stored rather than collecting these data points in a central location.

This raises questions about the allocation of data controller responsibility in a distributed processing environment. Additionally, it prompts considerations about the extent to which highly aggregated statistical analysis results can be considered anonymized. Guidelines are also needed for cryptographic protocols that enable blind secure aggregation of partial results (which themselves may need to be treated as encrypted personal data).

Specifically, we aim to explore these issues in relation to federated analysis, secure aggregation, and differential privacy. These techniques have been proposed as solutions to protect personal data against re-identification during distributed data processing.

The technical work in the proposed sandbox project will only handle synthetic data, although the ultimate goal of the SARDIN project is, of course, for the Health Data Bank to handle actual data containing personal information.

We base our approach on concrete examples of data-driven research and innovation from our project partners in the SARDIN project:

\begin{itemize}
    \item Region Östergötland: clinical research related to Parkinson’s disease.
    \item Region Västerbotten: stroke rehabilitation research.
    \item Dermacut: a company specializing in advanced wound image analysis.
\end{itemize}

The technical work in the proposed project will initially only handle strictly non-personal data synthesized from the project partners. That gives the opportunity to test workflows and prototypes before the questions of data controller responsibility are sorted. 
These practical examples will guide our efforts within the project. Of course, the long-term vision is to utilize actual health data with personally identifiable information in the Health Data Bank. 

To validate the proposed study, we conducted numerical experiments using four proxy datasets. These experiments aimed to test the proposed study's feasibility and effectiveness in a controlled environment. The results of these experiments are detailed in the Experiments and Results section.

\subsection{Contribution}
The ultimate goal is to enhance opportunities for researchers, innovators, and healthcare professionals to access data, including personal information, scale up projects, conduct world-leading research, and improve regional healthcare organizations' understanding of their operations and patients. This will be achieved through the development of the Health Data Bank platform.

Data protection and personal privacy are crucial foundational principles in modern information-driven and democratic economies. AI and other data-driven analytical tools have significant potential for future applications in precision medicine. However, these methods rely on healthcare systems effectively utilizing their data while adhering to privacy regulations. Our project aims to explore how cutting-edge privacy-enhancing technologies such as federated analysis, secure aggregation, and differential privacy can enable distributed data processing for these purposes within existing legislation.

Furthermore, we recognize that the questions we address within the project extend beyond a strictly healthcare context. Many sectors of society face similar trade-offs. We hope that the regulatory study we propose can also assist others in navigating this exciting landscape, which still lacks a comprehensive map.

In summary, our sandbox study aims to contribute to the development of guidelines and best practices that:

\begin{itemize}
    \item Assist data controllers, such as health care-providing regions, in protecting privacy while sharing analysis results.
    \item Support using EMR and clinical quality register for improved healthcare, research, innovation, and decision-making.
    \item Enable society to fully leverage the potential offered by securely analyzing and reusing health data held by various data controllers.
\end{itemize}

The conclusions from the sandbox study could also have implications for the use of privacy-enhancing technology beyond the healthcare domain.

\section{Background}

The General Data Protection Regulation (GDPR) is a European Union regulation that sets guidelines for collecting and processing personal data \cite{whatisgdpr}. As defined by the GDPR, personal data is any information relating to an identified or identifiable natural person \cite{whatispersonaldata}. This includes direct identifiers like names and identification numbers and indirect identifiers that could be used in combination with other data to identify an individual. The GDPR introduces the roles of data controllers and data processors \cite{whatisacontroller}. A data controller is a legal or natural person, an agency, a public authority, or any other body who, alone or jointly with others, determines the purposes and means of processing personal data. Under the GDPR, anonymization is the process of rendering personal data anonymous so that the data subject is no longer identifiable \cite{whatisanonymization}. This process must be irreversible, and once data is anonymized, it is no longer considered personal data under the GDPR.

The EU Artificial Intelligence (AI) Act represents a comprehensive legal framework for AI, with the objective of promoting the safe, trustworthy, and human-centric use of AI in the European Union \cite{AI_act_strategy,AI_act_briefing}. The AI Act categorizes AI systems into four risk levels: Unacceptable risk, High risk, Limited risk, and Minimal or no risk. Systems that pose an unacceptable risk are prohibited, while high-risk systems are subject to rigorous regulations. Regarding privacy protection, the AI Act operates in conjunction with the General Data Protection Regulation (GDPR). The AI Act, as a product safety law, governs the safe technical development and use of AI systems, while the GDPR, as a fundamental rights law, provides individuals with extensive rights concerning the processing of their data. The intention of this dual approach is to ensure that AI systems adhere to fundamental rights, safety, and ethical principles, as well as mitigate the risks associated with powerful and impactful AI models.

The European Health Data Space (EHDS) is a health-specific ecosystem comprised of rules, common standards and practices, infrastructures, and a governance framework \cite{ehds2024}. It aims to empower individuals through increased digital access to and control of their electronic personal health data at the national and EU levels. The intention is also to foster a single market for electronic health record systems, relevant medical devices, and high-risk AI systems. Furthermore, it identifies a trustworthy and efficient setup for the secondary use of health data for research, innovation, policy-making, and regulatory activities. Specifically, personal data can only be accessed and processed in closed, secure environments and only
anonymized data can be downloaded \cite{ehdsQaA}.

\subsection{Privacy enhancing technologies}

Privacy-enhancing technologies (PETs) can help organizations demonstrate a \textit{data protection by design and by default} approach to data processing, obtaining regulatory compliance, and protecting the individual's private data, which is an important ethical objective in itself. 

The Information Commissioner's Office (ICO) is the UK's independent authority for data protection and information rights, provides a helpful guide:   
The first part focuses on how PETs can help you achieve compliance with data protection law \cite{ICO2024HowCanPetsHelp_intro}.
The second part is intended for a more technical audience and briefly introduces eight types of PETs and explains their risks and benefits \cite{ICO2024WhatPetsAreThere_tech}.
In the paragraphs below, we revisit some important terms and refer the interested reader to the ICO report mentioned above for details. 

\textbf{Re-identification} refers to the post-processing of output data with the purpose of identifying individual persons or their personal information from source data. It is understood that output data for which the probability of re-identification significantly exceeds chance cannot be considered anonymized.

\textbf{Secure aggregation} is part of a family of cryptographic protocols that enables multiple parties to collectively compute the aggregation of their data without revealing it to each other (also known as multiparty computation, MPC). Even intermediate results are protected from other participants.

\textbf{Differential privacy} (DP) is a mathematical method that uses carefully calibrated noise to obscure an individual's contribution to a dataset, even when attackers can access other datasets. This technique allows researchers and database analysts to extract valuable information without disclosing personal identification details. Essentially, differential privacy ensures that the probability of a statistical query yielding a given result remains the same, regardless of whether the data includes an individual's data point. This randomness allows each individual to reasonably deny their membership in the dataset. Differential privacy can negatively impact the accuracy of the results. The goal is to strike a balance between preserving privacy and maintaining data utility for meaningful analysis.

\textbf{Homomorphic encryption} is a groundbreaking technology in the field of cryptography and cybersecurity that allows computations to be performed on encrypted data without the need for decryption\cite{FHE_Cheon2021,FHE_Hubaux2023}. 

\textbf{Decentralized Computation} refers to the processing of data across multiple locations without centralizing it. This approach enhances privacy by keeping data local and only sharing aggregated results or model updates. Federated Analysis often refers to statistical analysis over the distributed data.  
Federated Learning is a method where multiple parties jointly train a machine learning model without centrally collecting data (see, e.g., \ \cite{yang2019federated,lo2021systematic} for recent reviews).

\subsection{Regulatory sandboxes}

The European Union (EU) has been proactive in initiating regulatory sandboxes to examine legislation that concerns novel technology. The EU's Artificial Intelligence Act envisages setting up coordinated AI regulatory sandboxes to foster innovation in artificial intelligence (AI) across the EU \cite{EU_AI_act_regulatory_sandboxes}. These sandboxes intend to incentivize innovators to test their solutions in a controlled environment, allowing regulators to understand the technology better and foster consumer choice in the long run. 

Several U.S. state legislatures and regulators have explored the idea of regulatory sandbox programs, particularly for blockchain and fin-tech innovations. These sandbox initiatives vary significantly across states regarding their scope, the administering entity, and the participant pool \cite{newsJonesDay}.

Sandboxes operated by financial regulatory authorities are now widely used for fin-tech \cite{europarl_ai_study} as the practice is becoming more established also in other domains around the globe \cite{worldbank2020}. In the following paragraphs, we will highlight some regulatory sandbox exercises selected for their focus on advanced privacy-enhancing technologies. 

The ICO has been running the Regulatory Sandbox for several years\cite{ico_sandbox_previous_participants}.
The ICO Regulatory Sandbox provides a supportive environment for organizations creating innovative products and services that utilize personal data. One notable participant was Zamna, whose solutions leverage novel cryptography and distributed ledger technology for pre-airport checks while maintaining more robust passenger privacy levels. The study report was published end of 2023 \cite{ico_adewol2023zamna} 

The Infocomm Media Development Authority (IMDA) in Singapore has also established Privacy Enhancing Technology (PET) Sandboxes \cite{IMDA2024_PET_sandboxes} that are notable for considering cutting-edge technologies. 
This initiative aims to support businesses in piloting PET projects that address common business challenges while ensuring data privacy and protection. It has reported on four studies:

\begin{itemize}
    \item "Overcoming Data Barriers via Trustworthy Privacy-Enhancing Technologies" \cite{GPAI2023} conducted together with Global Partnership on AI (GPAI) examines how homomorphic encryption and differential privacy can be used to securely share and analyze data from the past pandemic.
    \item "Accessing more data through Trusted Execution Environment to generate new insights" \cite{imda_pet_sandbox_healthcare} examines the use of Trusted Execution Environment (TEE) by a pharmaceutical company to ensure that data from partners cannot be read by host of the environment while it is being pseudonymized. 
    \item "Preventing financial fraud across different jurisdictions with secure data collaborations" \cite{imda_pet_sandbox_mastercard} investigated the potential use of Homomorphic Encryption by Mastercard for sharing financial crime intelligence across international borders while complying with privacy legislation.
    \item "Digital advertising in a paradigm without 3rd party cookies" \cite{imda_pet_sandbox_meta} investigates how a combination of multiparty computing (MPC), aggregation, differential privacy (DP), and write-only identifiers could enable private measurements of digital ad attribution.
\end{itemize}

The Swedish Authority for Privacy Protection (IMY) has conducted sandbox studies since 2022.
IMY's approach follows a three-step process. First, they identify the legal issues that the guidance should focus on in collaboration with the participants. Then, they provide guidance over several months through workshops or other dialogue-based formats. Finally, they summarize their assessments in a public report to facilitate learning for a broader audience. To date, they have conducted the following projects and pilots:

\begin{itemize}
\item IMY conducted its first pilot \cite{imy_first_sandbox_pilot} with a regulatory sandbox in the autumn of 2022 in which two healthcare providers—Region Halland and Sahlgrenska University Hospital—collaborated to evaluate the possibilities of jointly training and exchanging machine learning models. The project, titled “Decentralized AI in Health Care: Federated machine learning between two healthcare providers,” aimed to better predict the re-admission of heart failure patients within 30 days of their last hospital stay. 

\item During the summer and autumn of 2023, IMY conducted its second pilot, which was titled "Safety Measurement in Public Environments Using IoT Technology" \cite{imy_third_sandbox_pilot}. Participants included the Traffic Office in the City of Stockholm, Internet of Things Sweden, and Kista Science City AB. The project aimed to use LiDAR sensors (a form of IoT sensing technology) to collect data on the proportion of women, men, and children visiting a public square. By understanding visitor demographics, the Traffic Office can enhance safety measures and promote the presence of underrepresented groups, such as women and children, in public spaces. Both data protection legislation and camera surveillance laws were considered in the study. 

\item A third sandbox study was conducted during the spring and summer of 2024 with Lidingö Municipality and Atea Sweden, which examined how generative AI can enhance the handling of public document disclosures by automating the identification and redaction of personal data. The study highlighted the importance of maintaining robust privacy protections, as the AI system effectively identified both direct and indirect personal data, minimizing the risk of human error and enhancing the accuracy of confidentiality assessments.

\item At this report's writing, a fourth IMY sandbox study was underway in collaboration with four major Swedish banks: SEB, Nordea, Swedbank, and Handelsbanken. This project aims to enhance information sharing to combat financial crime, such as fraud and money laundering while ensuring compliance with GDPR requirements. 
\end{itemize}

\subsection{Other related work}
The Swedish Medical Products Agency conducted a study focused on federated analyses \cite{Gedeborg2023article}. Their objective was to describe and facilitate the implementation of the methodology and technology of Federated Analyses. 
The report delves into statistical methods \cite{LV-rapport-teknisk}, legal considerations \cite{LV-rapport-rattsliga}, and IT implementation issues related to Federated Analyses, addressing both opportunities and challenges. It was developed in cooperation with statisticians from the Department of Medicine, Karolinska Institutet, Stockholm, and the Department of Surgical Sciences, Uppsala University.
IMY later released a consultation response commenting on the (limited) scope for shared controller responsibility \cite{IMY-remissvar-LV}. 

\section{Method}

\subsection{Proposed inquiry}

The project assumes that there is a large amount of health data containing individuals' personal information distributed across several healthcare regions, each being a data controller. This data is stored in isolated servers called edge nodes, see Figure \ref{fig:example}, which each is controlled by the healthcare region that also is the health data owner. We refer to this health data as source data and assume that it isn't necessarily evenly distributed across the edge nodes. Our goal is to calculate aggregated statistics over this source data, such as the mean value and standard deviation for a specific column.

A relevant question is whether such aggregated statistics can be used for re-identifying individuals and data subjects who have contributed data points, making them subject to data protection regulations. In practice, this risk depends on details within each individual dataset, including the number of data points, their distribution, and the intended data processing. Differential privacy has been proposed as a technical measure to control the risk of re-identification.

For our purposes, we further assume that:

\begin{enumerate}
    \item There is a legal basis for the local processing of personal data, allowing each edge node to compute a partial result based on its own source data,
    \item A statistical aggregate calculated over all data points in the distributed dataset cannot be used for re-identifying personal data; however,
    \item An aggregated partial result calculated from source data from a single edge node can still be associated with a risk of re-identification.
\end{enumerate}

For example, the total number of data points may be large enough to conceal each individual contribution to the aggregated final result. However, a single edge node may have only a few data points and cannot reveal its partial result without risking re-identification. Secure aggregation allows the edge nodes to collectively aggregate partial results under strong cryptographic guarantees, revealing in plaintext only the final result based on all health data

Our inquiries in the sandbox revolve around assessing whether the outcome, i.e., both intermediate and final results, of the described personal data processing constitutes personal data or not, as well as developing guidelines around the use of privacy-enhancing technologies for risk mitigation. The inquiries stated below aim to clarify the definition of anonymization in a distributed environment where PETs are implemented. 

\,

\noindent
Interpreting Anonymization and Aggregation:
\begin{itemize}
    \item Clarify the definition of anonymization and its purpose in data processing.
    \item Anonymization results from processing personal data to irreversibly prevent identification. Data controllers must consider all means reasonably likely to be used for identification.
    \item Anonymized data fall outside the scope of data protection legislation, but other provisions (e.g., confidentiality of communications) may still apply.
\end{itemize}

\noindent
Criteria for Aggregated Results to Be Considered Anonymous:
\begin{itemize}
    \item Define the criteria for aggregated results to be considered anonymous, such as the inability to identify individuals or link records to specific persons.
    \item Aggregated results should not allow individual identification or linking records to specific individuals, which may be the case, for example, for highly aggregated statistical results based on large data populations.
    \item Discuss solutions such as noise addition, differential privacy, statistical aggregation, secure multi-party aggregation, and homomorphic encryption.
\end{itemize}

\noindent
Distribution of Data Protection Responsibility in Federated Processing:
\begin{itemize}
    \item Clarify the concept of federated processing and its implications for data protection responsibility.
    \item The responsibility for anonymized aggregated results depends on the specific context and legal grounds.
    \item Data sharing agreements for distributed personal data handling need careful formulation.
    \item Analyse the separate but interconnected duties of the research principal and the data controller.
\end{itemize}

Furthermore, it would be interesting to investigate how the technical and organizational solutions we implement in the Health Data Bank align with the upcoming EHDS legislation, particularly in the context of Article 50 regarding the closed and secure processing of personal health data.

\subsection{Numerical Experiments}

To validate the proposed sandbox study, we conducted numerical experiments using four open-source de-identified datasets summarized in Table \ref{table:ds_info}: heart, framingham, adult, and brfss. These datasets are all popular benchmark datasets for machine-learning applications in public health and epidemiology

\subsubsection{Datasets.}
The four datasets used for the numerical experiments are:
\begin{itemize}
    \item \textbf{Heart}: The Cleveland Clinic Foundation Heart Disease Dataset is a well-known dataset frequently used in machine learning for heart disease prediction. It contains 14 attributes related to heart disease, such as age, sex, chest pain type, resting blood pressure, and serum cholesterol.
    \item \textbf{Framingham}: The Framingham Heart Study (FHS) dataset originates from a long-term cardiovascular study initiated in 1948 in Framingham, Massachusetts. It contains 16 attributes related to cardiovascular health.
    \item \textbf{Adult}: The Adult dataset, also known as the "Census Income" dataset, is widely used in machine learning for classification tasks. It contains 48,842 instances and 14 attributes, including age, workclass, education, marital status, occupation, relationship, race, sex, capital gain, capital loss, hours per week, and native country.
    \item \textbf{BRFSS}: The Behavioral Risk Factor Surveillance System (BRFSS) is a comprehensive health-related telephone survey system established by the Centers for Disease Control and Prevention (CDC). It gathers data on U.S. residents’ health-related risk behaviors, chronic health conditions, and use of preventive services.
\end{itemize}

\begin{table}
    \centering
    \small
    \caption{Summary statistics for the four open datasets used for the experiments listing the number of features (columns) and the number of rows. The original dataset (raw size) was cleaned of rows with missing values of duplicates (clean size). Balanced train and test set were then prepared for machine learning.\\}
    \begin{filecontents*}{info.txt}
dataset, features, raw size, clean size, train size, selected, predicted 
heart, 14, 1025, 302, 242, thalach, target
framingham, 16, 4240, 3658, 1114, totChol, TenYearCHD
adult, 7, 47592, 47592, 23078, capital-loss, income
brfss, 14, 464664, 413223, 41586, sleptim1, genhlth
\end{filecontents*}
\csvautobooktabular{info.txt}
    \label{table:ds_info}
\end{table}

\subsubsection{Experiments.}

To test the method we simulate three different scenarios: centralized processing with differential privacy, decentralized processing with local differential privacy, and secure decentralized processing with global differential privacy. Two main tasks are performed: comparing the means of a selected column for two subpopulations and training a logistic regression classifier to predict the target column. The selected feature column and predicted column for each dataset are listed in Table \ref{table:ds_info}, while other details are provided in Section \ref{sec:experiments}.

\subsection{Platform}

The Health Data Bank will handle the processing of personal data in a future real-world environment. This platform integrates several technical measures and processes to enhance privacy protection. In the proposed sandbox study, we will not directly process personal data. However, the study will investigate the processing of personal data as applied in the planned real-world environment.  

In simplified terms, the planned personal data processing in the Health Data Bank follows these steps:

\begin{enumerate}
    \item \textbf{Distributed Data Storage}:
        \begin{itemize}
            \item Source data is stored on edge nodes controlled by the regions. This is the original health data that contains personally identifiable information.
            \item This source data serves as the input for all subsequent data processing on the Health Data Bank platform.
        \end{itemize}
    \item \textbf{Distributed Data Processing}:
        \begin{itemize}
            \item Each edge node processes source data and produces partial results.
            \item Researchers send algorithms to the edge nodes via the Health Data Bank, allowing local processing without the source data leaving the edge node.
            \item Data processing is preceded by a sign-off from authorized personnel within the region's edge node.
            \item Intermediate results are generated during the distributed data processing, such as the output from the edge node's processing of source data.
            \item These partial results may be sensitive to re-identification and, despite being statistical aggregates, may need to be treated as personal data.
        \end{itemize}
    \item \textbf{Secure Aggregation}:
        \begin{itemize}
            \item Each edge node encrypts its partial result before sharing it, and aggregation occurs using a cryptographically secure protocol.
            \item Only the final result is decrypted and shared in plaintext with the researcher.
            \item Final (or partial) results can additionally be protected by adding DP noise.
            \item The final result is assumed to be highly aggregated non-personal data.        
        \end{itemize}
\end{enumerate}

\section{Numerical Experiments}
\label{sec:experiments}

\subsection{Experimental Setup}

We used four open-source datasets for the numerical experiments aiming to validate the proposed study method: heart, framingham, adult, and brfss, as discussed previously in relation to Table \ref{table:ds_info}. We assumed that the original data was evenly distributed over a network of edge nodes, where the size of the federation varied from $K=1$ to $K=64$ in powers of two.  

The experiments were conducted for a range of epsilon values from 0.01 to 100 with logarithmic stepping to identify critical values of the privacy parameters. Each trial was repeated at least 50 times for each target epsilon to collect sufficient samples for estimating confidence bands and assessing test accuracy. Two main tasks under three different scenarios were examined. 

\subsubsection{Tasks.}

Two main tasks were performed in the experiments:

\begin{itemize}
    \item \textbf{Task 1}: Comparing the means of a selected column for two subpopulations (split according to the target column for each dataset).
    \item \textbf{Task 2}: Training a logistic regression classifier to predict the target column for each dataset, using all other columns as input features.
\end{itemize}

\subsubsection{Scenarios.}

Each experiment simulated three different scenarios corresponding to different privacy assumptions about the secure processing environment:

\begin{itemize}
    \item \textbf{Scenario 1}: Centralized processing where all data is collected in a central repository and processed in plain text. Differential privacy is applied to protect the results, using the Laplace mechanism for t-stats and Renyi differential privacy for machine learning.
    \item \textbf{Scenario 2}: Decentralized processing where data remains at the local data node. Partial results (experiment 1) and gradients (experiment 2) are shared in plain text with a central server responsible for aggregation. Local differential privacy is applied, with each node in the federation adding noise to keep the partial results private from the server. The noise is tuned to the sensitivity and size of the local dataset.
    \item \textbf{Scenario 3}: Secure decentralized processing using a cryptographic protocol that keeps all partial results or gradients private from other parties. Global differential privacy is applied, with noise added based on the sensitivity and size of the entire distributed dataset, which is smaller than in scenario 2.
\end{itemize}

\subsubsection{Experiment 1: Federated Analysis.}

For task 1 we implemented a t-test to determine if the mean for the selected column was different for the two subpopulations. 
The distributed data processing to calculate the t-value becomes somewhat different for scenario 2 than for 1 and 3. Each node calculated the partial results, which are the mean and variance for the column for the local dataset. These now must be protected by dp noise sepqrately before they are shared with the server and aggregated in scenario 2. For scenario 1 and 3 it is only the final result, i.e., the t-stat, that is protected by the additive noise. 

The calculation for the mean is very straightforward, and we can use the Laplace mechanism right out of the box. However, as variance is strictly a non-negative quantity, adding noise according to an additive mechanism like Laplace or Gaussian could (and indeed will) sometimes result in negative values. To avoid this, we introduce the log-normal mechanism that adds Gaussian noise to the logarithm of the variance. According to the properties of the lognormal distribution, we now must use a bias correction term for the noise distribution, $\mathcal{N}(-0.5\sigma^2, \sigma^2)$, which is simply implemented for $\sigma$ which depends on the privacy parameters $\epsilon$ and $\delta$ for the conventional Gaussian mechanism. 

\subsubsection{Experiment 2: Federated Learning.}

For task 1 we trained a logistic regression model was trained for each data set under Renyí Differential Privacy, where data was either collected centrally (scenario 1) or spread out equally over a federation of different sizes (K). The maximal batch size was set to 242, and the learning rate was 0.01 using stochastic gradient descent with differential privacy. For the smaller dataset (Heart and Framingham), the effective batch size may have been smaller for large federations as each node becomes sparser.

\subsubsection{Data Processing.}

Each dataset was preprocessed to remove rows with missing values and duplicates. Columns were categorized as either numerical (values falling on a range) or categorical (values falling in discrete classes). Datasets were split into test and training sets in a 1 to 5 proportion relative to the minority class and balanced to ensure equal representation of target classes.

An exploratory feature analysis identified columns with the largest explanatory power for predicting the target value using a random forest classifier from the sklearn package in Python. This analysis informed the selection of columns for the t-test in experiment 1.

\subsection{Results}
\label{sec:results}

\begin{figure}
    \centering
    \begin{subfigure}{0.495\textwidth}
         \centering
         \includegraphics[width=\textwidth]{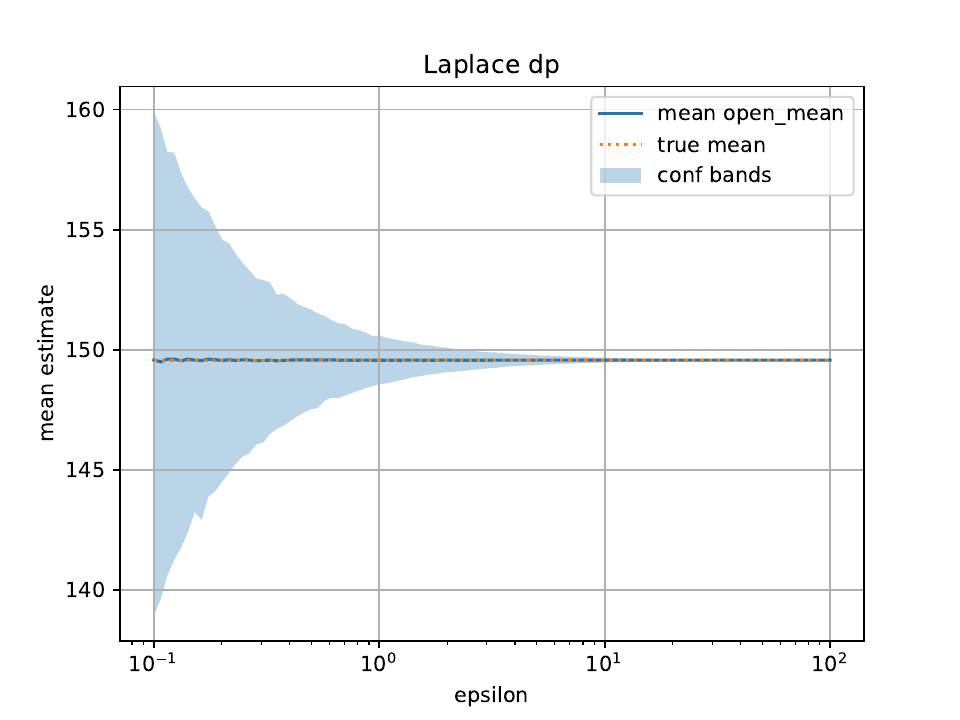}
         \caption{Uncertainty of the disclosed mean sampled by the Gaussian mechanism.}
         \label{fig:heart_laplace_thalach_mean_by_epsilon}
    \end{subfigure}
    \begin{subfigure}{0.495\textwidth}
         \centering
         \includegraphics[width=\textwidth]{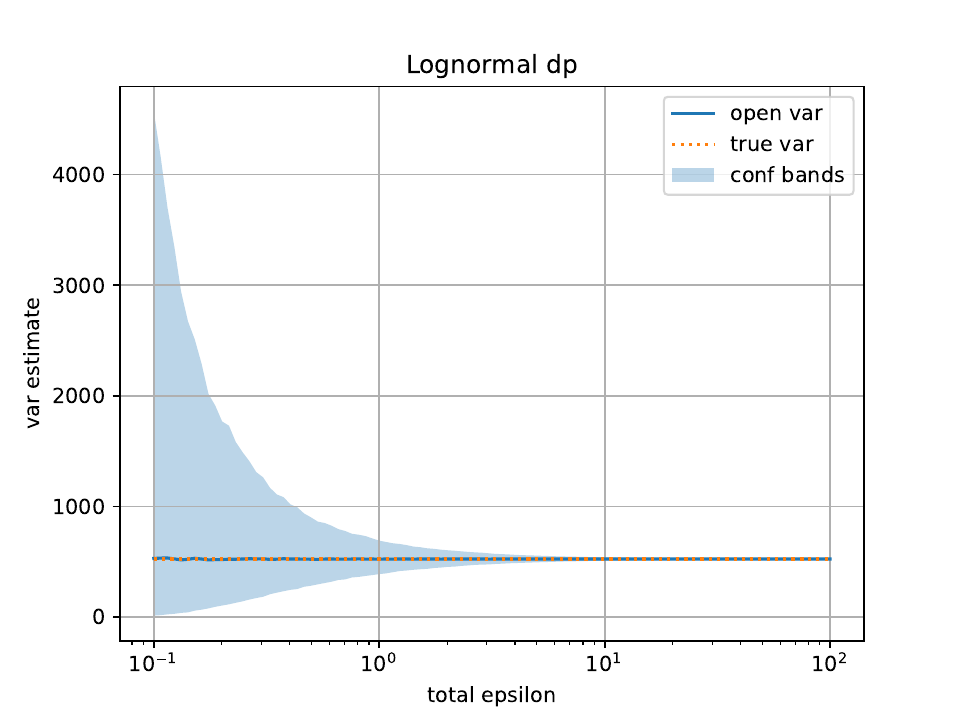}
         \caption{Uncertainty of the disclosed var sampled by the log-normal mechanism.}
         \label{fig:heart_lognormal_thalach_var_by_epsilon}
     \end{subfigure}
    \caption{Privacy-utlity trade-off for the thalach column of the entire heart dataset (assuming only one node, K=1) reported as the 95\% confidence band of disclosed value for 10000 samples at each value of the privacy parameter epsilon. We note from the right panel (b) that the sampled values for variance remain non-negative for the log-normal mechanism. }
    \label{fig:heart_thalach_central}
\end{figure}

\begin{figure}
    \centering
    \begin{subfigure}{0.495\textwidth}
        \centering
        \includegraphics[width=\textwidth]{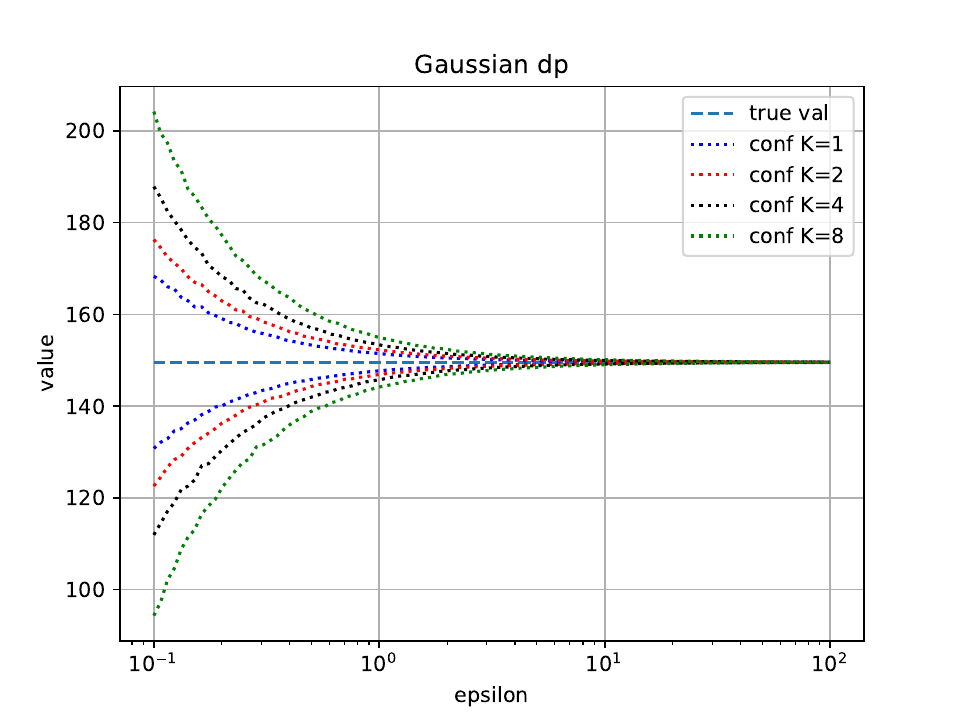}
        \caption{Uncertainty of the mean for different federation sizes (Gaussian mechanism)}
        \label{fig:heart_federated_gaussian_thalach_mean_by_epsilon}
    \end{subfigure}
    \begin{subfigure}{0.495\textwidth}
        \centering
        \includegraphics[width=\textwidth]{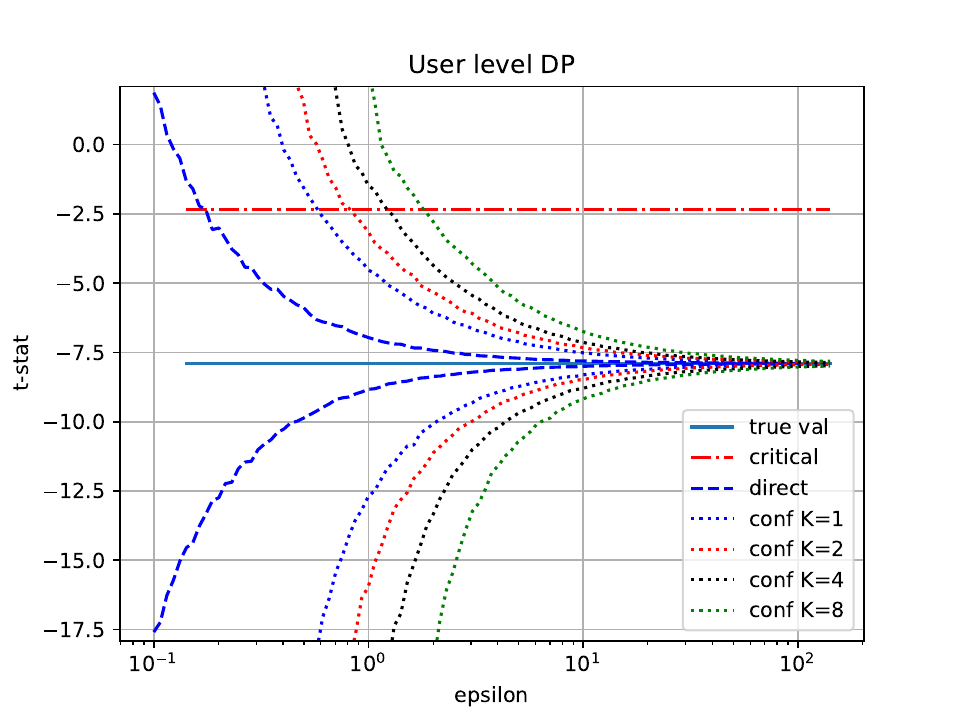}
        \caption{Uncertainty of the t-stat for different federation sizes (combined mechanism).}
        \label{fig:heart_federated_combined_thalach_tstat_by_epsilon}
        \end{subfigure}
    \caption{Privacy-utlity trade-off for the thalach column of heart dataset for different sizes of the federation (K) reported as the 95\% confidence band for 10000 samples at each value of the privacy parameter epsilon. The red dash-dotted line displayed the critical t-value for making meaningful decisions by the test -- its intersection with the upper confidence envelope of the disclosed t-stat yields for each K yields its critical epsilon value.}
    \label{fig:heart_thalach_federated}
\end{figure}

The results of the numerical experiments are summarized in Tables \ref{table:dp_stats} and \ref{table:dp_train}, and Figures \ref{fig:heart_thalach_central} and \ref{fig:heart_thalach_federated}. These results are intended to provide insights into the privacy-utility trade-offs for different scenarios and tasks.

\begin{table}
    \caption{Results for experiment 1, where the baseline row reports critical epsilon values for each dataset (* for selected column) for scenario 1 (centralized, K=1) and scenario 3 (secure distributed, arbitrary many nodes K). The other rows report values for different sizes of the federation (K) under plaintext aggregation (assuming data is evenly split between nodes). We note that scenario 3 permits an order of magnitude lower privacy parameters for larger federations.\\}
    \centering
    \small
    \begin{filecontents*}{dp_stats.txt}
trial,K,heart*,framingham*,adult*,brfss*
%trial,K,heart:thalach,framingham:totChol,adult:capital-loss,brfss:SLEPTIM1
baseline,-,0.175,0.169,0.005,0.017
federated,2,0.868,1.744,0.046,0.043
federated,4,1.230,2.650,0.066,0.061
federated,8,1.870,3.756,0.087,0.087
federated,16,3.047,6.122,0.123,0.115
federated,32,5.324,10.698,0.174,0.152
federated,64,7.293,20.046,0.247,0.215
\end{filecontents*}
\csvautobooktabular{dp_stats.txt}
    \label{table:dp_stats}
\end{table}

\subsubsection{Reported Metric: Critical Epsilon.}

Epsilon ($\epsilon$) is a key parameter in differential privacy that quantifies privacy loss. Smaller epsilon values provide stronger privacy guarantees but introduce more noise, reducing accuracy. Larger epsilon values reduce noise, improving accuracy but weakening privacy protections.

The experiments analyzed how varying epsilon affects the accuracy of statistical estimates and machine learning models. The critical epsilon is the value at which the trade-off between privacy and utility is optimal, ensuring data remains sufficiently accurate for analysis while providing acceptable privacy protection.

\subsubsection{Results for Experiment 1: Federated Analysis.}

The critical epsilon values for Experiment 1 are presented in Table \ref{table:dp_stats}. The baseline row reports values for centralized processing (Scenario 1) and secure distributed processing (Scenario 3). The other rows report values for different sizes of the federation under plaintext aggregation. Notably, Scenario 3 that applies secure aggregation for the federated analysis permits an order of magnitude lower differential privacy parameters for larger federations, indicating a significant improvement in privacy protection without compromising accuracy.

The privacy-utility trade-off for the thalach column of the heart dataset (assuming only one node, K=1) is illustrated in Figure \ref{fig:heart_thalach_central}. The left panel (a) shows the uncertainty of the disclosed mean sampled by the Gaussian mechanism, while the right panel (b) shows the uncertainty of the disclosed variance sampled by the log-normal mechanism. There is considerable uncertainty in the sampled values for the smaller epsilon, while for the largest epsilon, the sampled values are not visibly distinguishable from the true values. The panel (b) results indicate that the log-normal mechanism effectively maintains non-negative variance values, ensuring accurate statistical estimates.

Figure \ref{fig:heart_thalach_federated} shows the privacy-utility trade-off for the thalach column of the heart dataset for different sizes of the federation (K). The red dash-dotted line indicates the critical t-value for making meaningful decisions by the test -- its intersection with the upper confidence envelope of the disclosed t-stat yields the critical epsilon value for each K.

\begin{table}
    \centering
    \small
    \caption{Results for experiment 2, where the baseline row reports critical epsilon values for each dataset (selected column) for scenario 1 (centralized, K=1) and scenario 3 (secure aggregation, arbitrary many nodes K). The other rows report values for different sizes of the federation (K) under plaintext gradient aggregation (assuming data is evenly split between nodes). We note that scenario 3 (baseline) permits materially lower privacy parameters for larger federations.\\}
    \begin{filecontents*}{dp_train.txt}
trial,K,heart,framingham,adult,brfss
baseline,-,0.275,0.741,0.117,0.17
federated,2,0.417,1.07,0.155,0.191
federated,4,0.537,1.7,0.186,0.219
federated,8,1.29,1.86,0.331,0.363
\end{filecontents*}
\csvautobooktabular{dp_train.txt}
    \label{table:dp_train}
\end{table}

\subsubsection{Results for Experiment 2: Federated Learning.}

The critical epsilon values for Experiment 2 are reported in Table \ref{table:dp_train}. Recall that it corresponds to the epsilon for which the mean test set accuracy for the model trained with differential privacy is not statistically different from the non-private equivalent. 
In this experiment, we evaluated the impact of different federation sizes (K) on the critical epsilon values under various scenarios. The baseline row in the table presents the critical epsilon values for each dataset under two specific scenarios: scenario 1 that is the centralized computation ($K=1$) and scenario 3 that uses secure aggregation with and applies to a federation of arbitrary size (any $K\geq1$).

Centralized computation as in Scenario 1 has lower epsilon than any of the federated cases (Scenario 2) where the plaintext partial results have to be protected by noise injection at the node. The centralized scenario has the disadvantage that although the final results is protected, the data owners must share their data in plaintext with the central node.

In contrast, scenario 3 demonstrates the benefits of secure aggregation in a federated learning setup. Here, the critical epsilon values are indistinguishable from the centralized case for any size of the federation. This combines data privacy through the federated learning with an advantageous privacy-utility trade-off equivalent to the central case. Thus adding the secure aggregation protocol can enable larger federations to operate with materially lower privacy parameters while maintaining model accuracy.

\section{Discussion}

Although the first IMY sandbox study \cite{imy_first_sandbox_pilot} also treated distributed health data processing, the sandbox study proposal we present here is different in that we do not specifically focus on machine learning; we also explore conventional statistical methods as they are applied over a distributed dataset. We also examine other technical solutions, namely secure aggregation and differential privacy, which rely on strict assumptions and provide explicit technical guarantees for privacy protection. 
Furthermore, we assume that the aggregated end result, across all data from all nodes, can be considered anonymized, while any partial results (before global aggregation) may not be considered anonymized, for example, because the sample available at a local node is not sufficiently large. 

Gedeborg et al. at the Swedish Medical Product Agency \cite{Gedeborg2023article} also treated federated analysis of health data, although their investigation did not specifically consider using privacy enhancements like secure aggregation and differential privacy. Instead, they examined federated analysis where partial results were shared in plaintext with other participants in the network. Their analysis and models for shared controller responsibility among federated parties are, however, highly relevant to our project. 

Like secure aggregation and differential privacy, Federated analysis aims to facilitate efficient and accurate data processing while preserving privacy. It's essential to consider whether there are any material utility-privacy trade-offs in applying these PETs to the processing of personal data. 
A discussion of these trade-offs in relation to the proposed sandbox study follows below. 

\begin{table}
\centering
\caption{Different combinations of PET use in a hypothetical federated computation use-case, which assumes that partial results may be based on too few source data points to be considered non-personal data.\\}
\begin{tabular}{|c|c|c|}
\hline
\textbf{ } & \textbf{Differential privacy} & \textbf{No noise-injection} \\
\hline
\textbf{Secure aggregation} 
 & add global noise & final results disclosed \\
 & optimal accuracy (?) & full accuracy \\
 & tuneable risk (\textdagger) & reduced risk (*) \\
\hline
\textbf{Partial results open} 
 & add local noise & partial results disclosed \\
 & worse accuracy (?) & full accuracy \\
 & tuneable risk (\textdagger) & highest risk (*)\\
 \hline
\end{tabular}
\label{tab:PETcombos}
\end{table}

Table \ref{tab:PETcombos} outlines how using different PETs in a hypothetical federated computation use-case can result in different utility-privacy trade-offs. First, recall that we assume that partial results may be based on too few source data points to be considered non-personal data but that we assume the final aggregate result of the statistical analysis over the entire distributed data set can be regarded as non-personal data by GDPR.

The top row considers secure aggregation, where all partial results are private with very strong cryptographic guarantees, and we only need to consider disclosure of the final results, while for the bottom row, we assume that all partial results are (potentially) disclosed.
Conversely, the columns separate use cases where we use differential privacy (DP) to calibrate the risk of disclosing personal information about individual membership to an acceptable risk level. The accuracy under DP depends on the selected privacy parameter $\epsilon$, where the rightmost column represents infinite $\epsilon$, which corresponds to no differential privacy - note that final results here still can be considered anonymized in the GDPR sense for statistical aggregates.

We have indicated in the table by the symbol (\textdagger) that when we use differential privacy, the risk is tuneable as we can choose the parameters of the noise injection. We recall, however, from the empirical results for critical epsilon values in Section \ref{sec:results} that there are constraints on the noise injection parameters -- at too strong risk aversion, the accuracy suffers to the extent of rendering the output of the computations meaningless for the task at hand. This is indicated in the table by the symbol (?) to signal that it is uncertain whether one can achieve both acceptable accuracy and acceptable privacy risk in the DP-sense for a particular case. 
The symbol (*) in the table draws attention to the fact that without differential privacy with its tuneable parameters, the privacy risk is fixed and given by the size of the datasets. 
There are furthermore some technical details, particularly the secure aggregation without differential privacy case in the upper right quadrant of the table, which are further discussed in the third bullet in the list below. 

\begin{itemize}
    \item The top left quadrant of the table describes the case that uses both differential privacy and secure aggregation, as we propose for the Health Data Bank. Global noise that was calibrated to achieve acceptable (differential) privacy has been added to the results. Here, we assume that we can maintain an acceptable level of accuracy, that is, that the utility-privacy trade-off is relatively benign.
    This was indicated by the baseline row in Tables \ref{table:dp_stats} and \ref{table:dp_train}. 
    \item In the bottom left quadrant, on the other hand, we aggregate partial results in plain text, which requires the noise to be added at the local level. Local differential privacy can have a much more adverse trade-off and may result in poor accuracy at privacy risk levels that we can tolerate.
    This observation is supported by the results for rows marked federated in Tables \ref{table:dp_stats} and \ref{table:dp_train}.
    \item On the top right, we have only protection by the secure aggregation. This may be sufficient if we are comfortable that the final result of the analysis is based on sufficiently many data points and is of an aggregation type (e.g., mean) that has a low risk of disclosing personal data. This is, therefore, indicated in the table as reduced risk. We can expect near full accuracy as the secure aggregation techniques, albeit requiring fixed point number representation (quantization), generally do not impact numerical performance very much for well-calibrated parameters (but may nevertheless consume significant computing power).
    \item In the bottom left quadrant, we assume no use of secure aggregation or differential privacy. This is, therefore, the case with the highest privacy risk. We expect full accuracy as the aggregation is not subject to noise, neither from quantization nor DP noise injection.
\end{itemize}

As part of the proposed sandbox study, we want to carry out numerical experiments to populate tables like Table \ref{tab:PETcombos} with actual numbers taken from the three data-driven research and innovation use cases from our partners in the SARDIN project. We believe that such quantitative exercises that compare privacy risk for different cases can contribute to more nuanced data protection considerations beyond the solely binary risk perspective. 

\subsubsection{Limitations.}

The assumption of negligible risk of re-identification of the final result is used for convenience for the proposed study as it allows us to consider it anonymized. In a real-world scenario, the re-identification risk would have to be assessed, for example, through numerical experiments. 
%
Techniques such as secure aggregation and differential privacy can introduce significant computational overheads. This may affect the scalability and efficiency of the proposed methods, especially when dealing with large datasets or real-time data processing
The numerical experiments are conducted using a limited number of open-source datasets. While these datasets are popular benchmarks, they may not represent the full diversity of data encountered in real-world healthcare scenarios.

\section{Conclusions}

Which combination of PETs to use in order not to disclose the final or partial results of a distributed statistical analysis depends on how high privacy risk a data controller can tolerate. If we are uncertain about the re-identification risk of the final aggregate results, we can bound the privacy risk by introducing differential privacy at an $\epsilon$ privacy level that we are more comfortable with. Numerical experiments that estimate and table utility-privacy trade-offs for actual use cases can aid researchers, controllers, and regulators in reasoning around privacy risk, which is the precise intention of our proposed sandbox study. 

Estimating the critical epsilon can be very useful when balancing privacy and utility. It helps determine the point at which the noise added to protect privacy does not compromise the utility of the data. In practical applications, selecting epsilon below the critical point ensures that the data can be used effectively for analysis and decision-making while adhering to privacy requirements.

\begin{credits}
\subsubsection{\ackname} We extend our heartfelt thanks to Anneli Nöu, Kristina Andersson, and Camilla Evensson for their project contributions and insightful discussions.
Our gracious appreciation also to our research and innovation partners in the SARDIN project: Region Östergötland, Region Västerbotten, and the medical image analysis company Dermacut.
This study was funded by Vinnova, Sweden's Innovation Agency (DNR: 2023-01840).

\end{credits}
%
%
%
\bibliographystyle{splncs04}
\bibliography{main}

\begin{thebibliography}{10}
\providecommand{\url}[1]{\texttt{#1}}
\providecommand{\urlprefix}{URL }
\providecommand{\doi}[1]{https://doi.org/#1}

\bibitem{ico_adewol2023zamna}
Adewole, A.: Regulatory sandbox final report: Zamna - a summary of zamna’s participation in the ico’s regulatory sandbox. Tech. rep. (2023), \url{https://policycommons.net/artifacts/11168927/regulatory-sandbox-final-report/12047884/}, cID: 20.500.12592/zs7h8h4

\bibitem{FHE_Cheon2021}
Cheon, J.H., Costache, A., Moreno, R.C., Dai, W., Gama, N., Georgieva, M., Halevi, S., Kim, M., Kim, S., Laine, K., Polyakov, Y., Song, Y.: Introduction to Homomorphic Encryption and Schemes, pp. 3--28. Springer International Publishing, Cham (2021). \doi{10.1007/978-3-030-77287-1_1}, \url{https://doi.org/10.1007/978-3-030-77287-1_1}

\bibitem{EU_AI_act_regulatory_sandboxes}
{EPRS | European Parliamentary Research Service}: Artificial intelligence act and regulatory sandboxes. Tech. rep. (2022), \url{https://www.europarl.europa.eu/RegData/etudes/BRIE/2022/733544/EPRS_BRI%282022%29733544_EN.pdf}

\bibitem{AI_act_briefing}
{EPRS | European Parliamentary Research Service}: {Briefing EU Legislation in Progress: Artificial intelligence act}. \url{https://www.europarl.europa.eu/RegData/etudes/BRIE/2021/698792/EPRS_BRI%282021%29698792_EN.pdf} (2024), accessed: 2024-04-27

\bibitem{ehdsQaA}
{European Commission}: {Questions and answers - EU Health: European Health Data Space (EHDS)}. \url{https://ec.europa.eu/commission/presscorner/api/files/document/print/en/qanda_22_2712/QANDA_22_2712_EN.pdf} (2022), accessed: 2024-04-27

\bibitem{AI_act_strategy}
{European commission}: {AI Act}. \url{https://digital-strategy.ec.europa.eu/en/policies/regulatory-framework-ai} (2024), accessed: 2024-04-27

\bibitem{ehds2024}
{European commission}: {The European Health Data Space (EHDS)} (2024), \url{https://www.european-health-data-space.com/}, accessed: 2024-04-27

\bibitem{whatisacontroller}
{European commission}: {What is a data controller or a data processor?} \url{https://commission.europa.eu/law/law-topic/data-protection/reform/rules-business-and-organisations/obligations/controllerprocessor/what-data-controller-or-data-processor_en} (2024), accessed: 2024-04-27

\bibitem{whatisgdpr}
{European commission}: {What is GDPR?} \url{https://gdpr.eu/what-is-gdpr/} (2024), accessed: 2024-04-27

\bibitem{whatispersonaldata}
{European commission}: {What is personal data?} \url{https://commission.europa.eu/law/law-topic/data-protection/reform/what-personal-data_en} (2024), accessed: 2024-04-27

\bibitem{Gedeborg2023article}
Gedeborg, R., Igl, W., Svennblad, B., Wilén, P., Delcoigne, B., Michaëlsson, K., Ljung, R., Feltelius, N.: Federated analyses of multiple data sources in drug safety studies. Pharmacoepidemiology and Drug Safety  \textbf{32}(3),  279--286 (2023). \doi{https://doi.org/10.1002/pds.5587}, \url{https://onlinelibrary.wiley.com/doi/abs/10.1002/pds.5587}

\bibitem{GPAI2023}
{Global Partnership on AI}: {Overcoming Data Barriers Trustworthy Privacy-Enhancing Technologies}. Tech. rep., Global Partnership on AI (November 2023), \url{https://www.imda.gov.sg/-/media/imda/files/programme/pet-sandbox/overcoming-data-barriers-via-trustworthy-privacy-enhancing-technologies---demonstration-report.pdf}

\bibitem{FHE_Hubaux2023}
Hubaux, J.P.: Homomorphic Encryption, pp. 35--39. Springer Nature Switzerland, Cham (2023). \doi{10.1007/978-3-031-33386-6_8}, \url{https://doi.org/10.1007/978-3-031-33386-6_8}

\bibitem{imda_pet_sandbox_healthcare}
{Infocomm Media Development Authority (IMDA)}: Accessing more data through trusted execution environment to generate new insights. \url{https://www.imda.gov.sg/-/media/imda/files/programme/pet-sandbox/jan2024_imda-pet-sandbox-case-study-healthcare-services.pdf} (2024), last updated: January 2024

\bibitem{imda_pet_sandbox_mastercard}
{Infocomm Media Development Authority (IMDA)}: Preventing financial fraud across different jurisdictions with secure data collaborations. \url{https://www.imda.gov.sg/-/media/imda/files/programme/pet-sandbox/imda-pet-sandbox--case-study--mastercard.pdf} (2024), last updated: March 12, 2024

\bibitem{IMDA2024_PET_sandboxes}
{Infocomm Media Development Authority (IMDA)}: {Privacy Enhancing Technology Sandboxes}. \url{https://www.imda.gov.sg/how-we-can-help/data-innovation/privacy-enhancing-technology-sandboxes} (2024), accessed: 2024-04-27

\bibitem{imda_pet_sandbox_meta}
{Infocomm Media Development Authority (IMDA)}: Privacy enhancing technology sandboxes. \url{https://www.imda.gov.sg/-/media/imda/files/programme/pet-sandbox/imda-pet-sandbox--case-study--meta.pdf} (2024), last updated: March 12, 2024

\bibitem{whatisanonymization}
{Information Commissioner's Office}: {Anonymisation: managing data protection risk code of practice}. \url{https://ico.org.uk/media/about-the-ico/consultations/2619862/anonymisation-intro-and-first-chapter.pdf} (2024), accessed: 2024-04-27

\bibitem{ICO2024HowCanPetsHelp_intro}
{Information Commissioner's Office}: {How can PETs help with data protection compliance?} (2024), \url{https://ico.org.uk/for-organisations/uk-gdpr-guidance-and-resources/data-sharing/privacy-enhancing-technologies/how-can-pets-help-with-data-protection-compliance/}

\bibitem{ICO2024WhatPetsAreThere_tech}
{Information Commissioner's Office}: {What PETs are there?} (2024), \url{https://ico.org.uk/for-organisations/uk-gdpr-guidance-and-resources/data-sharing/privacy-enhancing-technologies/what-pets-are-there/}

\bibitem{ico_sandbox_previous_participants}
{Information Commissioner's Office (ICO)}: {Previous Participants in ICO Regulatory Sandbox}, \url{https://ico.org.uk/for-organisations/advice-and-services/regulatory-sandbox/previous-participants/}

\bibitem{newsJonesDay}
{Insights by Jones Day}: {EU and U.S. "Sandboxes" Are Groundbreaking Tools}, \url{https://www.jonesday.com/en/insights/2023/04/eu-and-us--sandboxes-are-groundbreaking-tools}, accessed: April 27, 2024

\bibitem{lo2021systematic}
Lo, S.K., Lu, Q., Wang, C., Paik, H.Y., Zhu, L.: A systematic literature review on federated machine learning: From a software engineering perspective. ACM Computing Survey  \textbf{54}(95),  1--39 (2021)

\bibitem{europarl_ai_study}
{Policy Department for Economic, Scientific and Quality of Life Policies Directorate-General for Internal Policies, EU}: Regulatory sandboxes and innovation hubs for fintech. Tech. rep. (2020), \url{https://www.europarl.europa.eu/RegData/etudes/STUD/2020/652752/IPOL_STU%282020%29652752\_EN.pdf}

\bibitem{rise2024sardin}
RISE: {SARDIN project} (2024), \url{https://www.ri.se/en/what-we-do/projects/sardin}, accessed: 2024-04-27

\bibitem{LV-rapport-rattsliga}
{Swedish Medical Products Agency}: Rapport federerade analyser - rättsliga överväganden (2021), \url{https://www.lakemedelsverket.se/490173/globalassets/dokument/regeringsuppdrag/rapport-federerade-analyser---rattsliga-overvaganden.pdf}, dnr: 4.3.1-2020-017988

\bibitem{LV-rapport-teknisk}
{Swedish Medical Products Agency}: Rapport federerade analyser - teknisk beskrivning (2021), \url{https://www.lakemedelsverket.se/490889/globalassets/dokument/regeringsuppdrag/rapport-federerade-analyser-teknisk-beskrivning.pdf}, dnr: 4.3.1-2020-017988

\bibitem{IMY-remissvar-LV}
{The Swedish Authority for Privacy Protection (IMY)}: Remissvar läkemedelsverkets rapport federerade analyser (2021), \url{https://www.imy.se/globalassets/dokument/remissvar/2021/remissvar-lakemedelsverkets-rapport-federerade-analyser.pdf}

\bibitem{imy_third_sandbox_pilot}
{The Swedish Authority for Privacy Protection (IMY)}: Generativ ai är i fokus i vårens regulatoriska sandlåda (2024), \url{https://www.imy.se/nyheter/generativ-ai-ar-i-fokus-i-varens-regulatoriska-sandlada/}

\bibitem{imy2024sandbox}
{The Swedish Authority for Privacy Protection (IMY)}: Regulatorisk sandlåda om dataskydd | imy (2024), \url{https://www.imy.se/verksamhet/dataskydd/innovationsportalen/delta-i-var-regulatoriska-sandlada-om-dataskydd/}, accessed: 2024-04-27

\bibitem{imy_first_sandbox_pilot}
{The Swedish Authority for Privacy Protection (IMY)}: First regulatory sandbox pilot. Tech. rep. (Year of publication), \url{https://www.imy.se/globalassets/dokument/ovrigt/first-regulatory-sandbox-pilot---english-summary.pdf}

\bibitem{worldbank2020}
{World Bank}: Global experiences from regulatory sandboxes (2020), \url{https://documents1.worldbank.org/curated/en/912001605241080935/pdf/Global-Experiences-from-Regulatory-Sandboxes.pdf}, accessed: April 27, 2024

\bibitem{yang2019federated}
Yang, Q., Liu, Y., Chen, T., Tong, Y.: Federated machine learning: Concept and applications. ACM Transactions on Intelligent Systems and Technology (TIST)  \textbf{10}(2), ~12 (2019)

\end{thebibliography}

\end{document}